\documentclass[superscriptaddress,amssymb,amsmath]{revtex4}

 \usepackage{graphics}

 \usepackage{amssymb}
 \usepackage{amsthm}
 \usepackage{amsmath}
 \usepackage{mathtools}
 \usepackage{natbib}

\usepackage{ifpdf}
\usepackage{subfigure}
\usepackage{psfrag}
\usepackage{graphicx}

\newcommand{\zip}[1]{ }

\def\@fnsymbol#1{\ifcase#1\or *\or \dagger\or \ddagger\or \mathchar "278\or \mathchar "27B\or \|\or **\or \dagger\dagger \or \ddagger\ddagger \else\@ctrerr\fi\relax}
\long\def\symbolfootnote[#1]#2{\begingroup%
\def\thefootnote{\fnsymbol{footnote}}\footnote[#1]{#2}\endgroup}

\begin{document}

\title{Energy harvesting using vortex-induced vibrations of tensioned cables}
\author{Cl\'ement~Grouthier}
\email{clement.grouthier@ladhyx.polytechnique.fr}
\affiliation{LadHyX, Department of Mechanics, Ecole Polytechnique, 91128 Palaiseau, France}
\author{S\'ebastien~Michelin}
\email{sebastien.michelin@ladhyx.polytechnique.fr}
\affiliation{LadHyX, Department of Mechanics, Ecole Polytechnique, 91128 Palaiseau, France}
\author{Emmanuel~de~Langre}
\email{delangre@ladhyx.polytechnique.fr}
\affiliation{LadHyX, Department of Mechanics, Ecole Polytechnique, 91128 Palaiseau, France}
\date{\today}

\begin{abstract}
The development of energy harvesting systems based on fluid/structure interactions is part of the global search for innovative tools to produce renewable energy. In this paper, the possibility to harvest energy from a flow using vortex-induced vibrations (VIV) of a tensioned flexible cable is analyzed. The fluid loading on the vibrating solid and resulting dynamics are computed using an appropriate wake-oscillator model, allowing one to perform a systematic parametric study of the efficiency. The generic case of an elastically-mounted rigid cylinder is first investigated, before considering an infinite cable with two different types of energy harvesting : a uniformly spanwise distributed harvesting and then a periodic distribution of discrete harvesting devices. The maximum harvesting efficiency is of the same order for each configuration and is always reached when the solid body and its wake are in a frequency lock-in state. 
\end{abstract}


\maketitle

\section{Introduction}

Geophysical flows such as wind, oceanic currents or tides represent a clean source of energy that is widely available. Many devices exist, or are being imagined, to harvest part of this energy, ranging from wind farms to wave energy converters or marine turbines, to cite a few. 

Among all these, a specific class is based on flow-induced vibrations of structures. Such vibrations may be caused by several distinct mechanisms, see for instance reviews in \citet{Blev} or \citet{Naud}. Fluid elastic instabilities, a major mechanism, have been considered in the perspective of energy harvesting by \citet{Tang}, \citet{Barr}, \citet{Doar} or \citet{Sing}, on plates, cylinders or squares. Vortex-induced vibrations (VIV) is another important cause of flow-induced vibrations and originates in a strong dynamical coupling between an oscillating body and its fluctuating wake, causing lock-in of frequencies and high-amplitude motion \citep{Will}. \citet{Bern} proposed an energy harvesting device using VIV and the idea has since then been further developed, \citep{Yosh,Barr2}.

A key question for all energy harvesting devices is how to access large quantities of energy : because of the low energy density in geophysical flows, large systems are required. In the specific domain of VIV, such large systems have actually been extensively studied for offshore engineering issues such as mooring cables or risers, see for instance \citet{Baar}. This opens new perspectives in harvesting energy, using VIV of very long tensioned flexible structures instead of elastically supported rigid short cylinders. 

Although the local interaction mechanisms between each body section and its wake are still the same, a new dimension is introduced when considering the spanwise behaviour of such slender structures. This results in a much more complex dynamics, involving vortex-induced waves, travelling or stationary. These have been observed on a variety of systems \citep{Alex,Facc2,Vand,Moda}. Extensive numerical computations \citep{Newm,Willd,Luco,Bour} have shown rich dynamics for this configuration of a cable coupled with its wake. A simpler approach, using a wake oscillator model, recently allowed to derive some generic results on complex issues such as mode switching, time sharing and space sharing in complex flows \citep[][and references within]{Viol2}. This approach has been extensively compared with experiments and DNS computations \citep{Facc2,Viol,Viol2} and used for predictive analysis \citep{Xu,Srin}.

In the present paper we seek to establish some fundamental results on generic configurations of energy harvesting from VIV of tensioned cables. To this end, the wake oscillator model introduced in \citet{Facc} is used, as well as its linearized version \citep{Lang}. In Section~\ref{section2D}, the elementary 2D case of an elastically supported rigid cylinder is considered. The infinite tensioned cable with continuously distributed harvesters is solved in Section~\ref{section3Dcontinu}. Finally in Section~\ref{section3Dperiodique} the influence of the harvesters distribution is addressed. The general features of this new idea of harvesting energy from vortex-induced waves in cables are then discussed in Section~\ref{sectiondiscussion}.
	
\section{The elastically mounted rigid cylinder}
\label{section2D}

\subsection{Fluid-solid model}

\begin{figure}[!ht]
	\centering
	\psfrag{a}[cc][cc][0.65]{(a)}
	\psfrag{b}[cc][cc][0.65]{(b)}
	\psfrag{c}[cc][cc][0.65]{(c)}
	\psfrag{r}[cc][cc][0.75]{$r$}
	\psfrag{h3}[cc][cc][0.65]{$h$}
	\psfrag{xi3}[cc][cc][0.65]{$r$}
	\psfrag{xi}[cc][cc][0.65]{$r$}
	\psfrag{Xi}[cc][cc][0.65]{$R$}
	\psfrag{Theta}[cc][cc][0.65]{$\Theta$}
	\psfrag{Theta2}[cc][cc][0.65]{$\Theta$}
	\psfrag{L}[cc][cc][0.65]{$L$}
	\psfrag{D}[cc][cc][0.65]{$D$}
	\psfrag{U}[bc][cc][0.75]{$U$}
	\psfrag{x}[cc][cc][0.75]{$x$}
	\psfrag{y}[cc][cc][0.75]{$y$}
	\psfrag{z}[cc][cc][0.75]{$z$}
	\includegraphics[width = 0.3\textwidth]{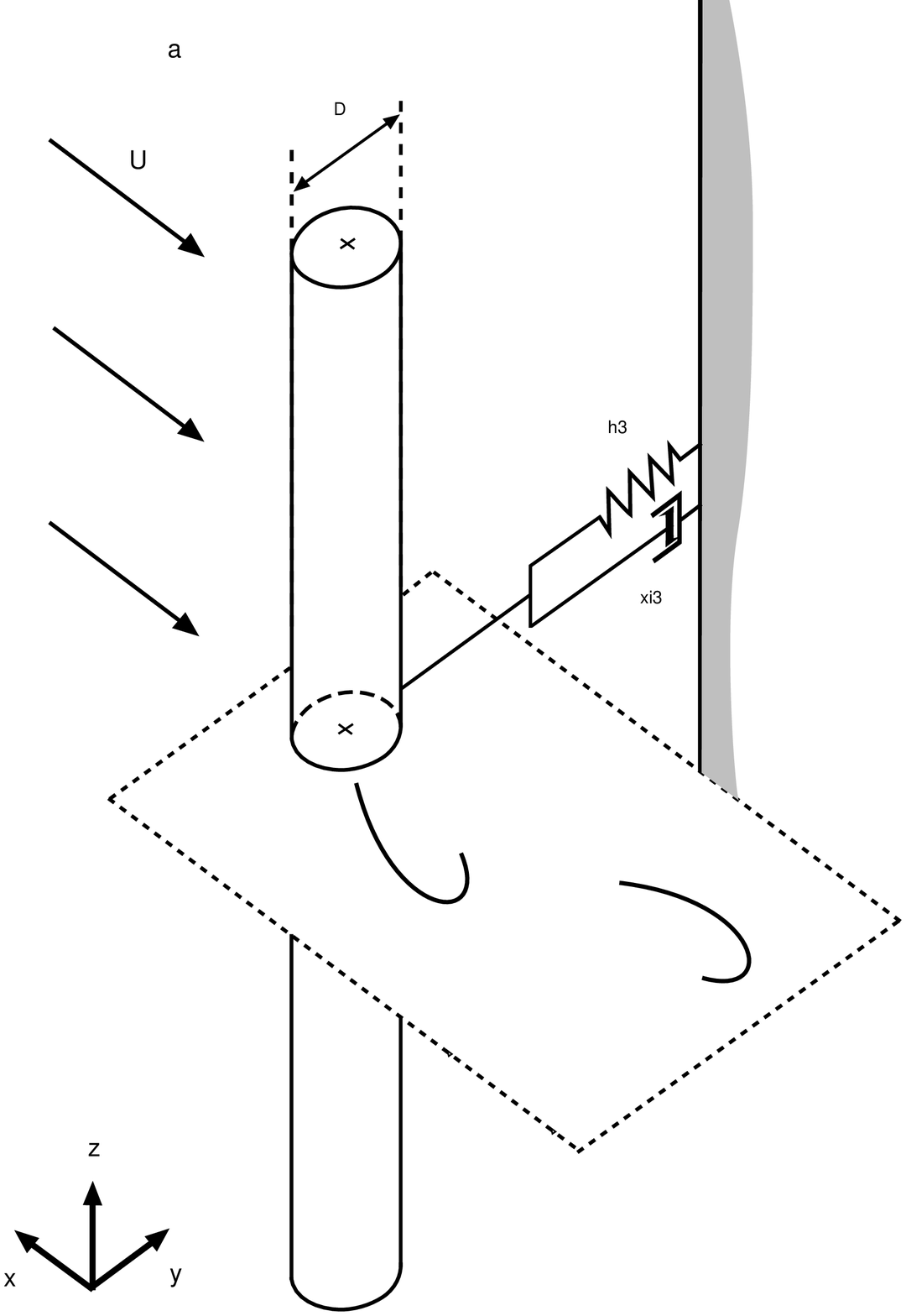} \hfill
	\includegraphics[width = 0.3\textwidth]{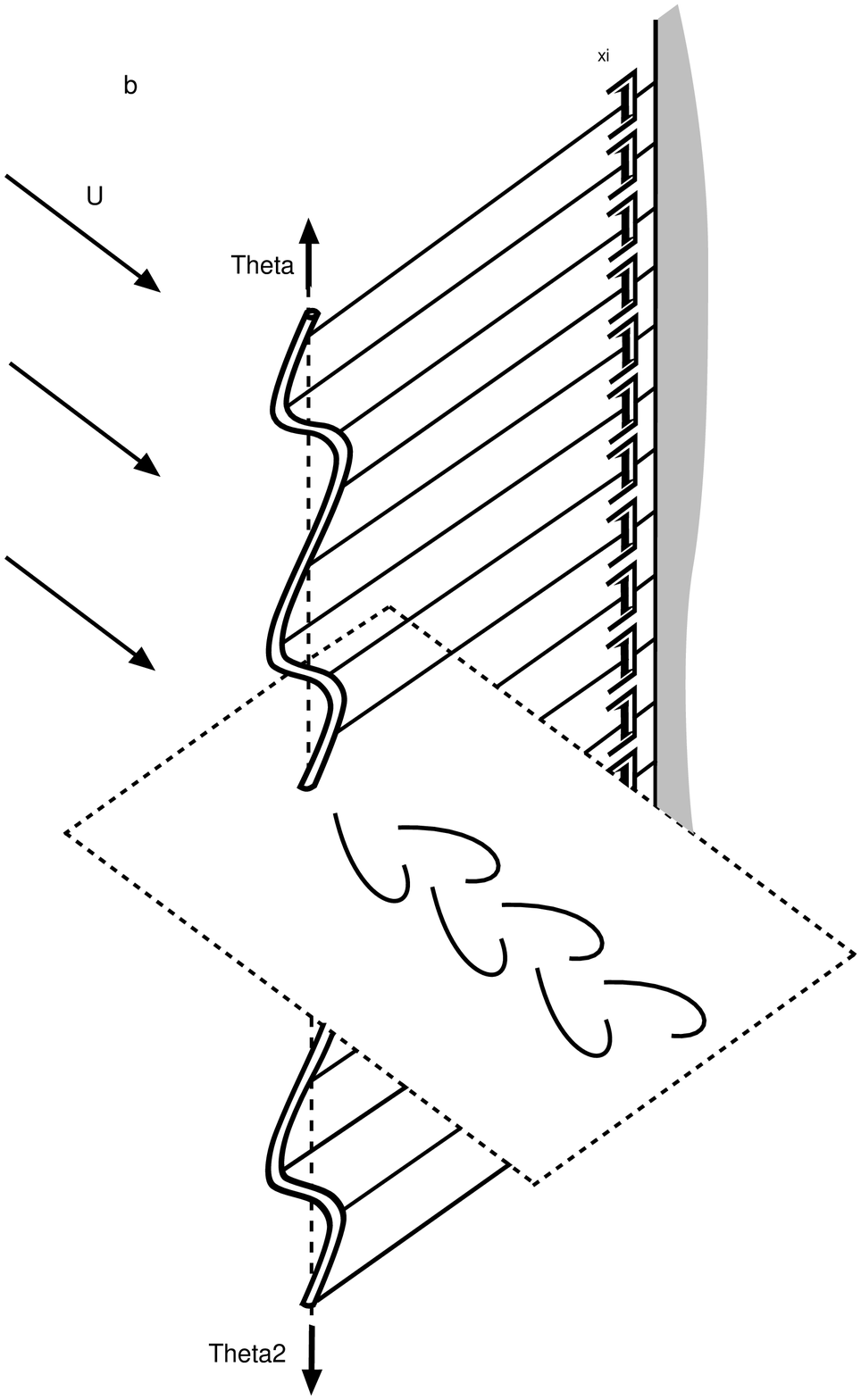} \hfill
	\includegraphics[width = 0.3\textwidth]{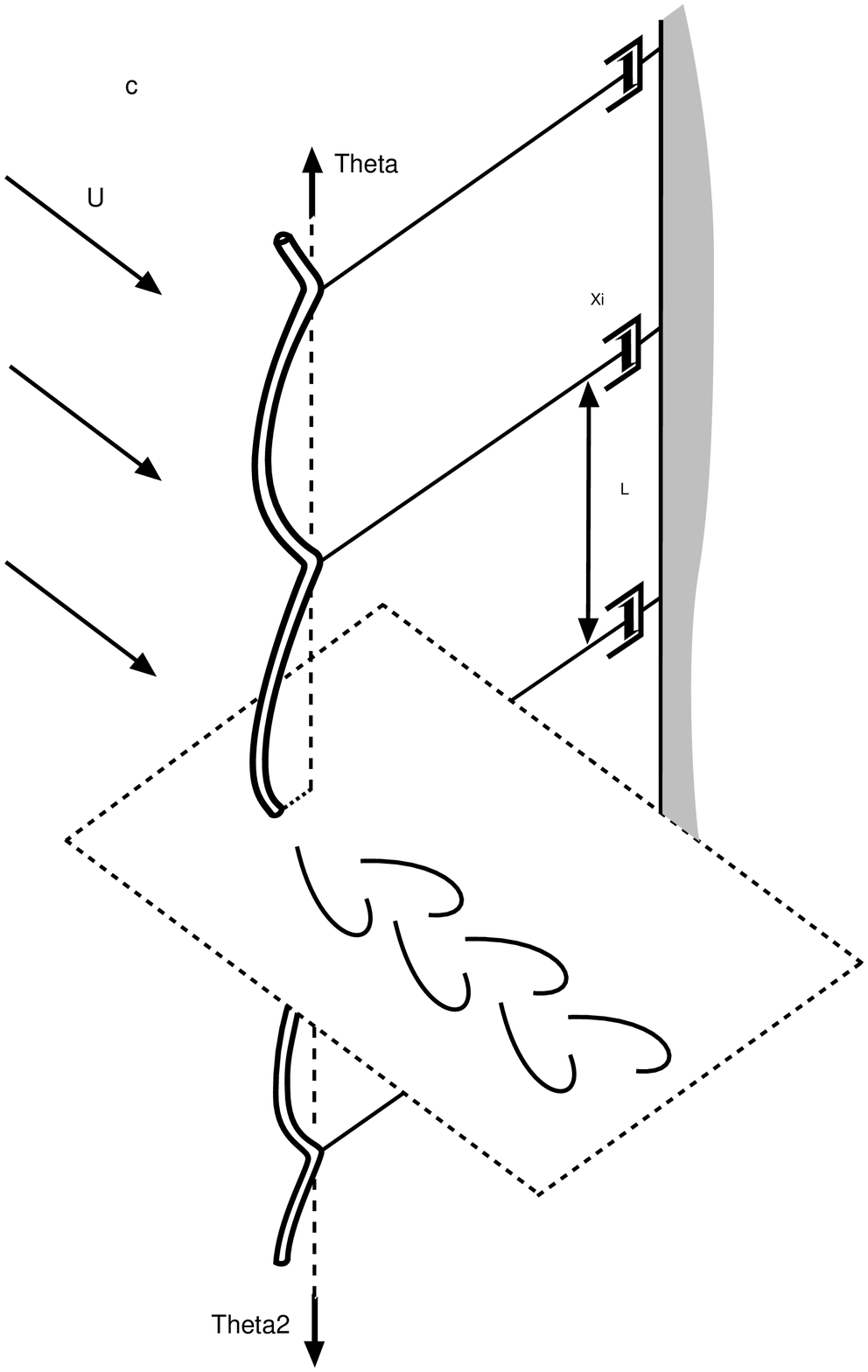} \hfill
	\vspace{0.5cm}
	\caption{Energy harvesting from the vortex-induced vibrations of (a) the 2D generic case of an elastically mounted rigid cylinder, (b) an infinite tensioned cable with continuous energy harvesters, (c) an infinite tensioned cable with localized energy harvesters.}
	\label{schema} 
\end{figure}

We consider first the classical generic two dimensional case (2D) of a cylinder mounted on an elastic and damped support, Figure \ref{schema}(a). Let $r$ and $h$ be the damping and stiffness coefficients of the support, $D$, $m_{s}$ and $Y$ the diameter, mass per unit length and cross-flow displacement of the cylinder and $\rho$, $U$ the fluid density and velocity. Following previous authors \citep{Blev,Facc}, the cross-flow dynamics of the cylinder under flow is described by 
\begin{equation}
	\left( m_{s} + \dfrac{\pi}{4} \rho D^{2} C_{M0} \right) \dfrac{\partial ^{2} Y}{\partial T^{2}} + \left( r + \dfrac{1}{2} \rho D U C_{D} \right) \dfrac{\partial Y}{\partial T} + h Y = F_{wake},
	\label{eq_solide_2d}
\end{equation}
where the fluid loading has been split into three parts : the added mass with a coefficient $C_{M0}$, the added damping with the drag coefficient $C_{D}$ and the remaining part denoting the wake force $F_{wake}$.

Two frequencies exist in the system : (i) that of the cylinder in still fluid ($U=0$), namely $\omega_{s} = \sqrt{ h/ m_{t}}$, where $m_{t} = m_{s} + \rho \pi D^{2} C_{M0} /4$ is the total mass per unit length, and (ii) the vortex shedding frequency behind the still cylinder, $\omega_{f} = 2 \pi St U/D$, where $St$ is the Strouhal number \citep{Blev}. Using the latter, the dimensionless form of \eqref{eq_solide_2d} reads 
\begin{equation}
	\ddot{y} + \left( \xi + \dfrac{\gamma}{\mu} \right) \dot{y} + \delta^{2} y = f_{wake},
	\label{eq_solide_ad}
\end{equation}
where $y = Y/D$, $t = \omega_{f} T$, $\delta = \omega_{s}/\omega_{f}$ and $\dot{\left( \mbox{ } \right)}$ denotes derivation with respect to the dimensionless time $t$. The dimensionless damping is composed of the support damping $\xi = r/m_{t} \omega_{f}$ and of the fluid damping $\gamma/ \mu$ where $\gamma = C_{D}/ 4 \pi St$ is the stall parameter and $\mu = m_{t} / \rho D^{2}$ is the mass ratio \citep{Skop2,Facc}. 

From the point of view of the fluid-solid system, energy harvesting induces a loss of energy, which will be represented in the remainder of the paper by the support damping $\xi$. The efficiency of the harvesting may then be defined as in \citet{Barr,Barr2} or \citet{Zhu}.
\begin{equation}
	\eta = \dfrac{ \left\langle r \left( \dfrac{\partial Y}{\partial T} \right)^{2} \right\rangle}{\frac{1}{2} \rho D U^{3}},
	\label{efficiency_formula}
\end{equation}
where $ \left\langle \mbox{ } \right\rangle $ denotes averaging in time. Using dimensionless variables, this also reads 
\begin{equation}
	\eta = 16 \mu \pi^{3} St^{3} \left\langle  \xi \dot{y}^{2} \right\rangle.
	\label{efficiency_formula_ad}
\end{equation}
	
The wake force, $f_{wake}$ in equation \eqref{eq_solide_ad}, is now modelled using the wake oscillator dynamics proposed in \citet{Facc} and used for example in \citet{Viol,Viol2} or \citet{Xu}. The fluctuating lift coefficient $q = 2 C_{L} / C_{L0}$ is assumed to satisfy a Van der Pol equation forced by the cylinder acceleration, so that the coupled cylinder and wake equations read respectively
\begin{subequations} \label{ad_model_2D}
	\begin{gather}
	\ddot{y} + \left( \xi + \dfrac{\gamma}{\mu} \right) \dot{y} + \delta^{2} y = Mq, \label{ad_model_2D_1} \\
	\ddot{q} + \varepsilon \left( q^{2}-1 \right) \dot{q} + q = A \ddot{y}, \label{ad_model_2D_2}
	\end{gather}
\end{subequations}
where $M = C_{L0}/16 \mu \pi^{2} St^{2}$ and $A$, $\varepsilon$ are two dimensionless parameters based on experimental data. In all the paper, we shall take $A=12$, $\varepsilon = 0.3$, $C_{D} = 2$, $C_{L0} = 0.8$, $St = 0.17$, $\mu = 2.79$ and $C_{M0} = 1$ as in previous work \citep{Viol2}, so that $M = 0.06$ and $\gamma / \mu = 0.34$ in equations \eqref{ad_model_2D}.

Equations \eqref{ad_model_2D} are integrated in time using finite differences. Using initial conditions with the cylinder at rest and the wake variable subject to a small random perturbation, the system reaches a limit cycle within a few periods. This cycle is then analyzed in terms of frequency and time-averaged quantities, such as the efficiency, equation \eqref{efficiency_formula_ad}.

\subsection{Optimal energy harvesting and lock-in}

\begin{figure}[!ht]
	\centering
	\psfrag{x0}[cc][cr][0.65]{$10^{-2}$}
	\psfrag{x1}[cc][cr][0.65]{$10^{-1}$}
	\psfrag{x2}[cc][cc][0.65]{$1$}
	\psfrag{x3}[cc][cc][0.65]{$10$}
	\psfrag{y1}[cc][cc][0.65]{$1$}
	\psfrag{y2}[cc][cc][0.65]{$2$}
	\psfrag{y3}[cc][cc][0.65]{$3$}
	\psfrag{c1}[lc][cc][0.65]{$0.1$}
	\psfrag{c2}[lc][cc][0.65]{$0.2$}		
	\psfrag{sigmadelta}[tc][bc][0.9][0]{$\xi$}
	\psfrag{delta}[cc][cc][0.9][-90]{$\delta$}
	\psfrag{a}[cc][cc][0.65][0]{(a)}		
	\includegraphics[width = 0.45\linewidth]{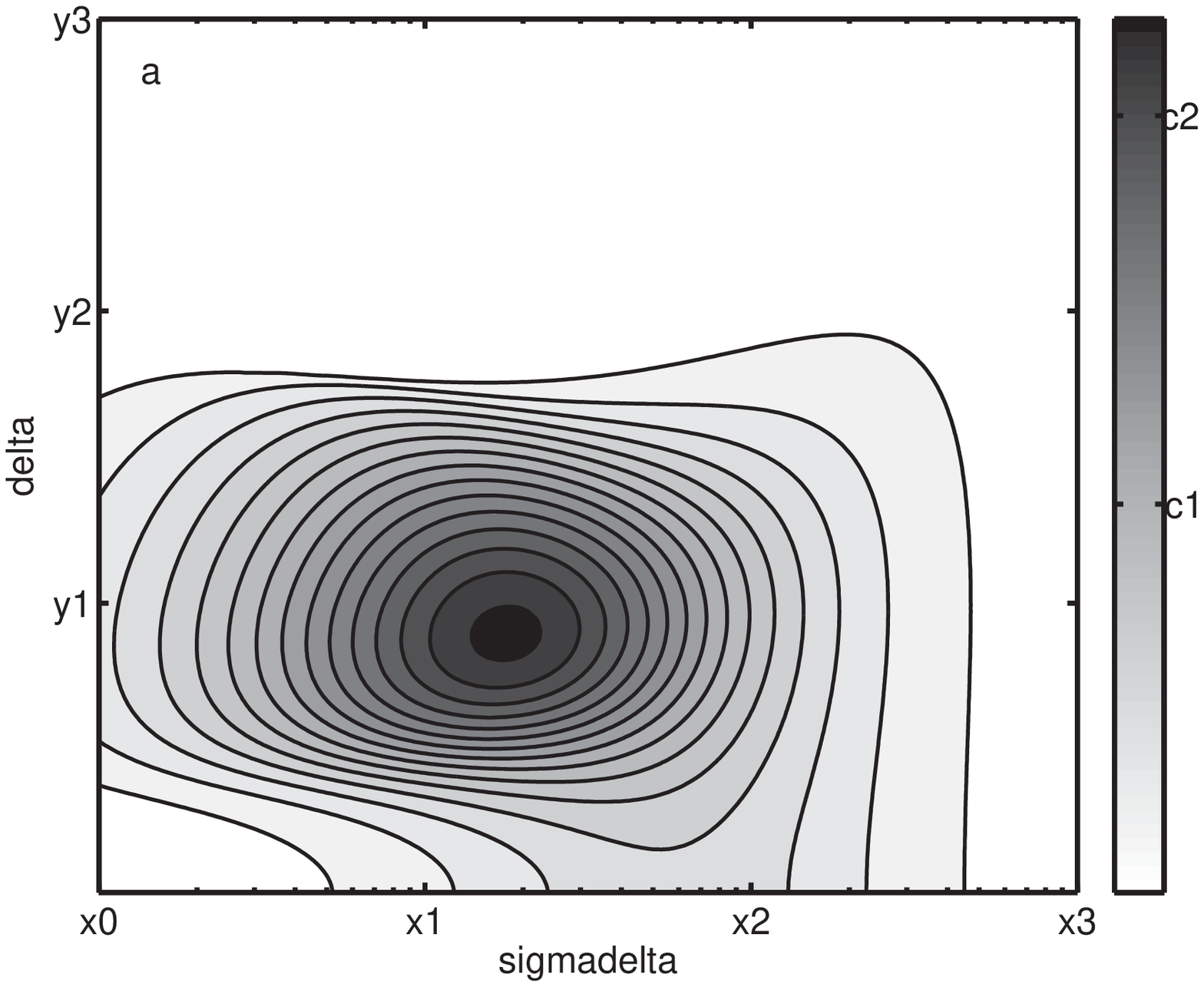} \hfill
	\psfrag{c1}[lc][cc][0.65]{$0.2$}
	\psfrag{c2}[lc][cc][0.65]{$0.4$}
	\psfrag{c3}[lc][cc][0.65]{$0.45$}		
	\psfrag{b}[cc][cc][0.65][0]{(b)}		
	\includegraphics[width = 0.45\linewidth]{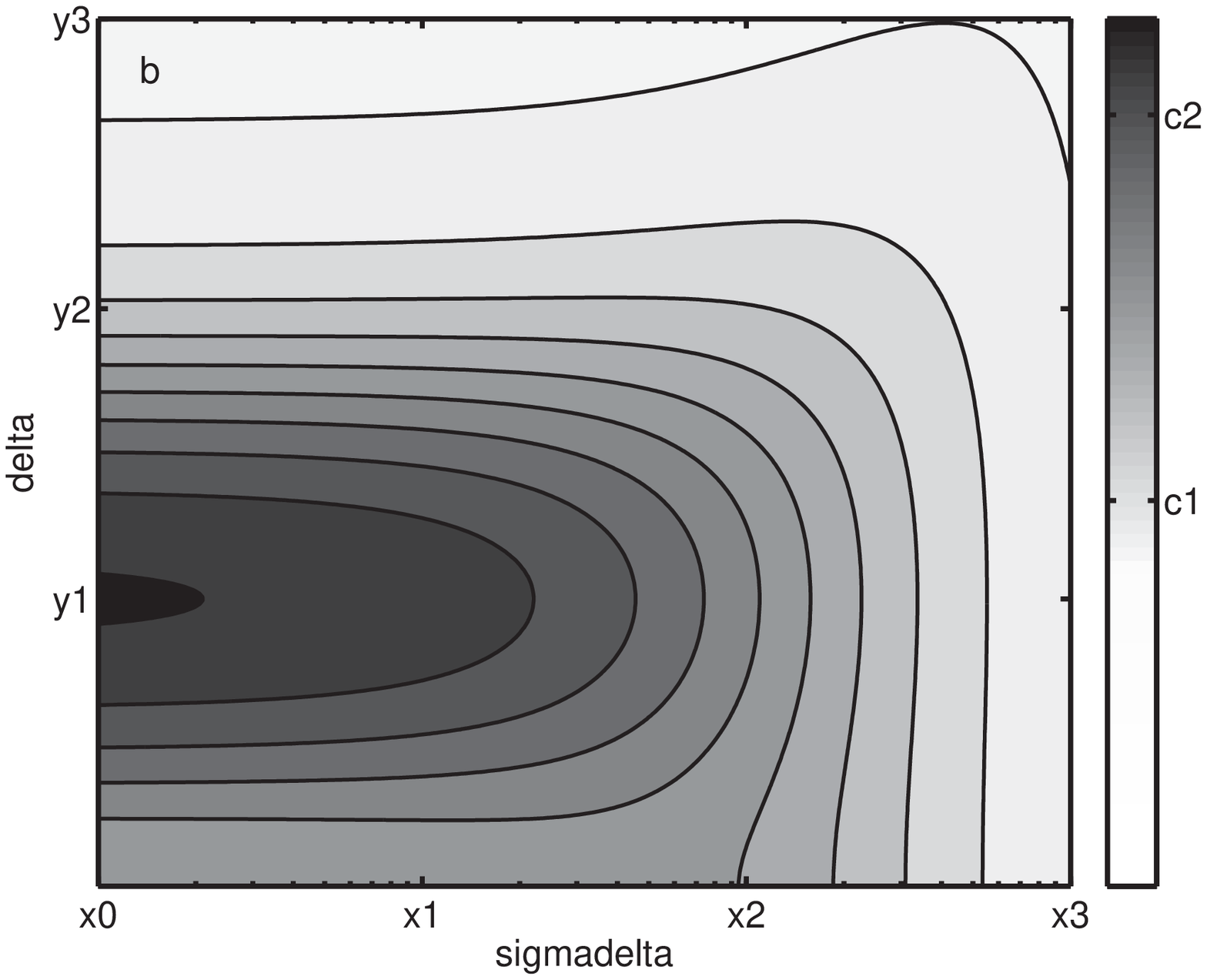}
	\vspace{0.5cm}
	\caption{Effect of the damping $\xi$ and the frequency ratio $\delta$ in the 2D model : (a) Harvesting efficiency using the full non-linear model (level step : 0.015), (b) Growth rate derived from the linear model (level step : 0.03)}
	\label{efficiency_2d} 
\end{figure}

In the present case, the harvesting efficiency $\eta$ depends only on the frequency ratio $\delta$ and the reduced damping $\xi$. Figure \ref{efficiency_2d}(a) in fact shows a very strong dependence of the efficiency with the parameters, notably the frequency ratio $\delta$. The optimal efficiency $\eta = 0.23$ is reached for $\delta = 0.89$ and $\xi = 0.20$. This is consistent with the value of $\eta = 0.22$ obtained experimentally in \citet{Bern}. The model allows one to recover an expected result which is that the efficiency vanishes for both small and large dampings. In the limit of small damping, the amplitude of VIV saturates, as is known from the Skop-Griffin plot \citep{Will}, and the efficiency then varies linearly with $\xi$ and tends to zero. For high dampings, the amplitude decreases like $1/\xi$ so that the efficiency, which scales as $\xi \dot{y}^{2}$, will also vanish. In terms of frequency ratio between the solid and fluid frequency $\delta = \omega_{s}/\omega_{f}$, the optimal is close to one. This corresponds to the classical lock-in condition, known to lead to high amplitudes. 

To model lock-in, a simplified version of the wake oscillator model was proposed in \citet{Lang}. Linearizing equations \eqref{ad_model_2D}, a linear stability analysis of the system was then amenable : lock-in appeared to correspond to the highest growth rate of the coupled mode instability involving the cylinder and the wake. In the present case, the linear equations read
\begin{subequations} \label{lin_model_2D}
	\begin{gather}
	\ddot{y} + \left( \xi + \dfrac{\gamma}{\mu} \right) \dot{y} + \delta^{2} y = Mq, \label{lin_model_2D_1}\\
	\ddot{q} - \varepsilon \dot{q} + q = A \ddot{y}. \label{lin_model_2D_2}
	\end{gather}
\end{subequations}
Note that we keep here all the linear terms, including damping, whereas only the dominant terms were used in \citet{Lang}. Looking for harmonic solutions in time leads to the frequency equation for $\omega$
\begin{equation}
\omega^{4} + i\omega^{3} \left( \varepsilon - \alpha \right) + \omega^{2} \left( \varepsilon \alpha + AM - 1 -\delta^{2} \right) + i \omega \left( \alpha - \varepsilon \delta^{2} \right) + \delta^{2} = 0,
\label{polynome_2D}
\end{equation}
where $\alpha = \xi + \gamma/\mu$. The dynamics of the system will be dominated by the most unstable mode \citep{Lang,Viol2}. Its growth rate is shown on Figure \ref{efficiency_2d}(b) as a function of $\xi$ and $\delta$. Figure \ref{efficiency_2d}(b) displays both the lock-in effect for $\delta = 1$ and the decrease of the growth rate for large values of $\xi$, two features that were found on the efficiency map (Figure \ref{efficiency_2d}a). In terms of efficiency, large values are found to lie in the zone of large linear growth rates. 

This section shows that a simple VIV model with damping allows to capture the main expected effect of the two harvesting parameters, the frequency ratio and the intensity of damping/harvesting. The optimal efficiency is reached for a clear lock-in condition and a well balanced damping value.

\section{Infinite cable with distributed energy harvesting}
\label{section3Dcontinu}

We consider now an infinite tensioned cable as a three-dimensional generalization of the elastically mounted rigid cylinder of the previous section. Again, the harvesting device is modelled by a dissipation introduced in the system, here a uniformly distributed damping support (Figure \ref{schema}b).

The dynamics of the cable and of the wake are modelled as in the previous section, but with the noticeable change that the stiffness term $hY$ in equation \eqref{eq_solide_2d} is now replaced by a tension induced stiffness $- \Theta \partial^{2} Y / \partial Z^{2}$, where $\Theta$ is the uniform tension in the cable and $Z$ the spanwise coordinate \citep{Facc2,Viol}. The corresponding dimensionless equations are 
\begin{subequations}\label{ad_model_continu}
	\begin{gather}
	\ddot{y} + \left( \xi + \dfrac{\gamma}{\mu} \right) \dot{y} - y^{\prime \prime} = Mq, \label{ad_model_continu_1} \\
	\ddot{q} + \varepsilon \left( q^{2}-1 \right) \dot{q} + q = A \ddot{y}, \label{ad_model_continu_2}
	\end{gather}
\end{subequations}
where $z = Z \omega_{f} \sqrt{m_{t}/\Theta}$ and the derivation with respect to the dimensionless spanwise coordinate $z$ is noted $\left( \mbox{ } \right)^{\prime}$, all other variables and parameters being identical to those of the previous section. Note that we use here a set of dimensionless variables that differs from the above cited papers, for the sake of further clarity. These equations are now integrated in time and space using finite differences on a spatially periodic domain. Its size is tested to be long enough compared with the wavelength of the developing motion for it not to influence the result. Initially, the cable is at rest and a small random perturbation is applied to the wake variable $q$. Only the steady state response is here considered. 

\begin{figure}[!ht]
	\centering
	\psfrag{x0}[cc][bc][0.65]{$-1$}
	\psfrag{x1}[cc][bc][0.65]{$0$}
	\psfrag{x2}[cc][bc][0.65]{$1$}
	\psfrag{y0}[cc][cc][0.65]{$0$}
	\psfrag{y1}[cc][lc][0.65]{$2.5$}
	\psfrag{y2}[cc][lc][0.65]{$5$}	
	\psfrag{y}[tc][bc][0.9][0]{$y(z,t)$}
	\psfrag{z}[cc][lc][0.9][-90]{$\dfrac{z}{2 \pi}$}	
	\psfrag{a}[cc][cc][0.65][0]{(a)}	
	\includegraphics[width = 0.45\linewidth]{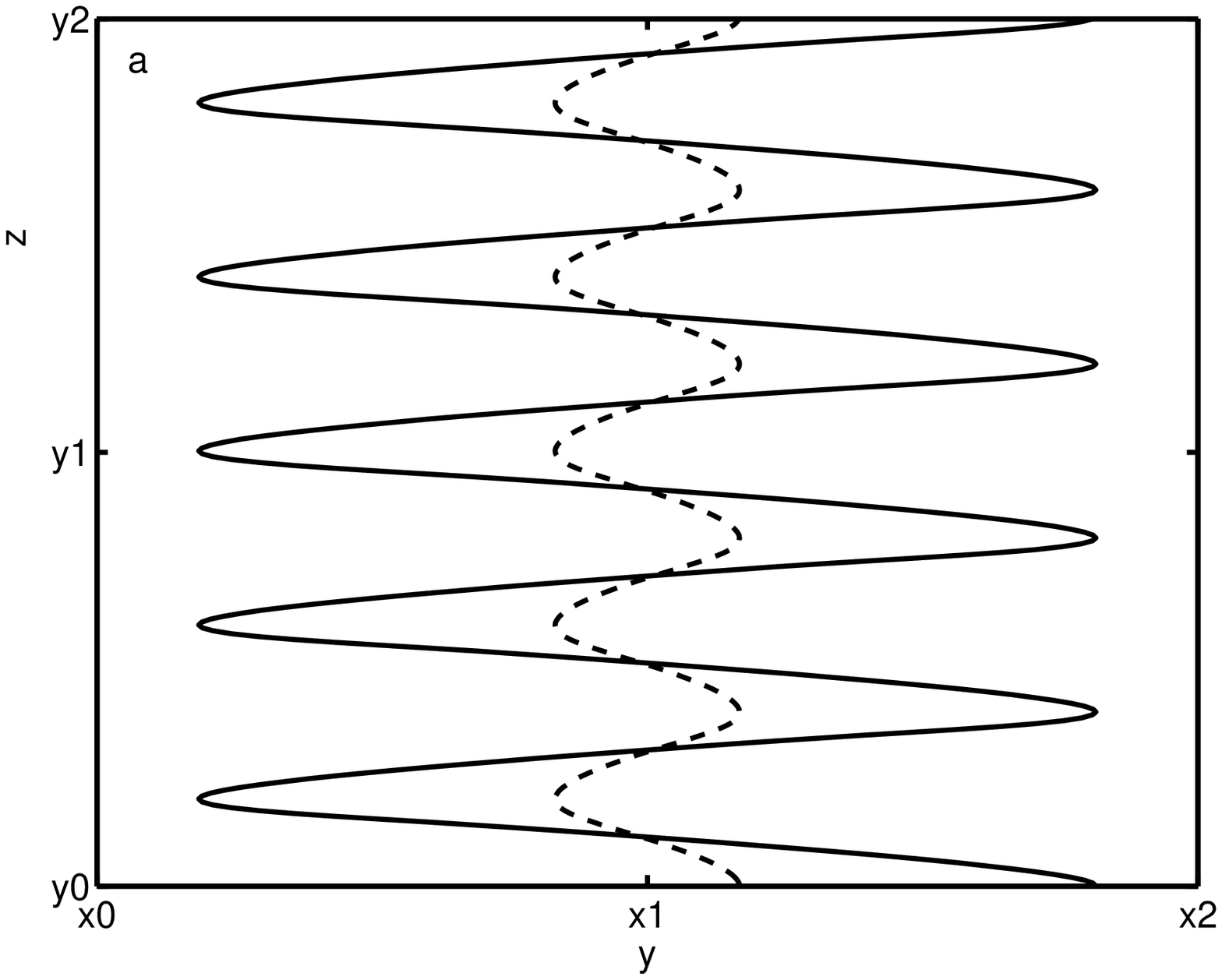}	\hfill
	\psfrag{x0}[cc][br][0.65]{$10^{-2}$}
	\psfrag{x1}[cc][br][0.65]{$10^{-1}$}
	\psfrag{x2}[cc][bc][0.65]{$1$}
	\psfrag{x3}[cc][bc][0.65]{$10$}
	\psfrag{y1}[cc][cc][0.65]{$0$}
	\psfrag{y2}[cc][lc][0.65]{$0.1$}
	\psfrag{y3}[cc][lc][0.65]{$0.2$}	
	\psfrag{sigmadelta}[tc][bc][0.9][0]{$\xi$}
	\psfrag{eta}[cc][cc][0.9][-90]{$\eta$}
	\psfrag{b}[cc][cc][0.65][0]{(b)}		
	\includegraphics[width = 0.45\linewidth]{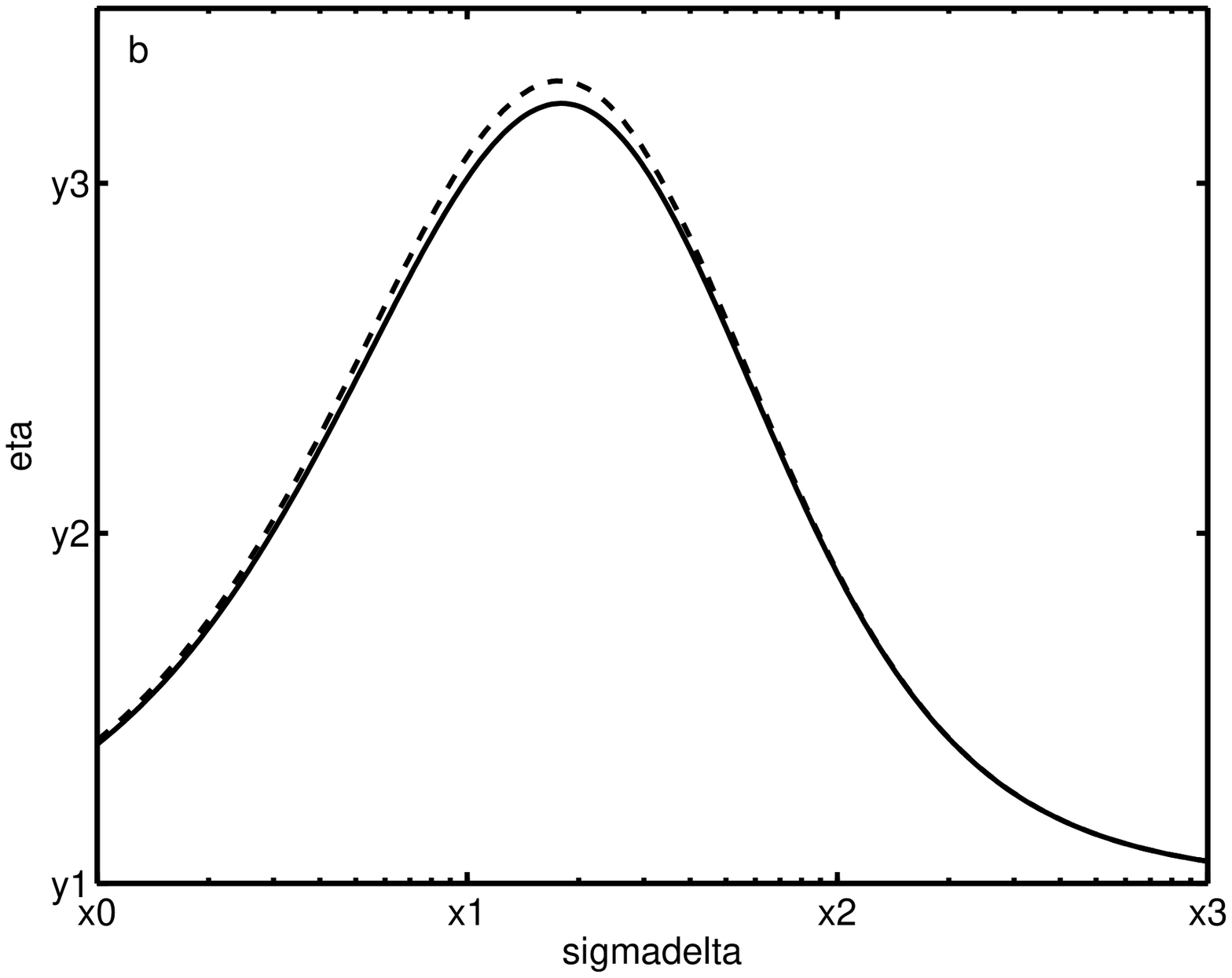}
	\vspace{0.5cm}
	\caption{(a) Instantaneous displacement of the cable for different values of the damping for $\xi = 0.1$ (solid) and $\xi = 1$ (dashed). (b) Comparison of the optimal harvesting efficiency $\eta_{2D} \left( \xi \right)$ from the elastically mounted rigid cylinder (dashed) and that of the infinite tensioned cable with distributed harvesters, $\eta_{3D} \left( \xi \right)$ (solid).}
	\label{efficiency_continu} 
\end{figure}

The influence of the only harvesting parameter, the damping $\xi$, is now explored. It is found that the system responds in the form of propagating quasi-harmonic waves, as previously described in \citet{Facc2}, \citet{Viol} and illustrated on Figure \ref{efficiency_continu}(a). Note that these waves have an amplitude that depends on $\xi$ but their wavelength apparently does not. The efficiency is uniform in span and varies as a function of $\xi$ in the classical bell shape (Figure \ref{efficiency_continu}b). An optimal is found, $\eta = 0.22$ for $\xi = 0.18$. These values are very close to those found in the 2D case, but not identical.

To better understand these dynamics, the linearized equations are used, as previously. These read here
\begin{subequations}\label{lin_model_continu}
	\begin{gather}
	\ddot{y} + \left( \xi + \dfrac{\gamma}{\mu} \right) \dot{y} - y^{\prime \prime} = Mq, \label{lin_model_continu_1} \\
	\ddot{q} - \varepsilon \dot{q} + q = A \ddot{y}. \label{lin_model_continu_2}
	\end{gather}
\end{subequations}
Looking for a harmonic solution both in space and time, the dispersion relation between the complex frequency $\omega$ and the real wavenumber $k$ is found as
\begin{equation}
\omega^{4} + i\omega^{3} \left( \varepsilon - \alpha \right) + \omega^{2} \left( \varepsilon \alpha + AM - 1 -k^{2} \right) + i \omega \left( \alpha - \varepsilon k^{2} \right) + k^{2} = 0,
\label{polynome_continu}
\end{equation}
where $\alpha = \xi + \gamma/\mu$. This equation is identical to that obtained in the 2D case, equation \eqref{polynome_2D}, but for the replacement of the frequency ratio $\delta = \omega_{s}/\omega_{f}$ by the wavenumber $k$. We know from the previous section that the highest growth rate is obtained for $\delta = 1$, hence $k = 1$ here. This explains two features of the dynamics of the cable. First, the numerically-observed wavenumbers are actually within $10^{-4}$ of this reference value $k=1$ for all explored values of $\xi$ (Figure \ref{efficiency_continu}a) : the system does always select the linearly dominant wavenumber. This is consistent with the work of Violette \citep{Viol2}. As a result, the 3D efficiency curve $\eta_{3D} \left( \xi \right)$ (Figure \ref{efficiency_continu}b) is strictly equivalent to the cut of the efficiency map of the elastically-mounted rigid cylinder at the level $\delta = 1$ (Figure \ref{efficiency_2d}a). For each damping $\xi$, the optimal frequency ratio on Figure \ref{efficiency_2d}(a) differs slightly from $\delta = 1$. An efficiency curve $\eta_{2D} \left( \xi \right)$ is obtained from the map Figure \ref{efficiency_2d}(a) by selecting the optimal $\delta$ for each damping. It is plotted on Figure \ref{efficiency_continu}(b) : the 3D and 2D curves are very close but not identical. In particular, the 3D optimal efficiency $\eta = 0.22$ slightly differs from the 2D optimum $\eta = 0.23$.

It is interesting to note that even if the tensioned cable case is not exactly as efficient as the elastically supported cylinder one, it has a major advantage in the fact that it adapts its stiffness by adapting its wavelength so that it is always at lock-in. A rigid cylinder on the other hand only has a short range of operability in terms of flow velocity, which corresponds to $\omega_{s} = \omega_{f}$. This will be further discussed in Section~\ref{sectiondiscussion}.

\section{Infinite cable with localized energy harvesting}
\label{section3Dperiodique}

\subsection{Fluid-solid model with discrete energy harvesting}

In practice, most energy extractors are not continuous devices, as assumed in the previous section. We now address the case where the corresponding dampers are periodically spaced along the length of the cable (Figure \ref{schema}c). All the features of the model are identical to those of the preceding section, except dissipation, which is now governed by the the dynamics of each damper,
\begin{equation}
	\Theta \left[ \dfrac{\partial Y}{\partial Z} \right]^{+}_{-} = R \dfrac{\partial Y}{\partial T},
	\label{BC}
\end{equation}
where $\left[ \mbox{ } \right]^{+}_{-}$ stands for the jump between values on both sides of a damper and $R$ is the damping coefficient. Here $L$ is the distance between two consecutive dampers. We focus on cable motions that are periodic in space over a length $L$, so that the dimensionless equations reduce to
\begin{subequations}\label{ad_model_periodique}
	\begin{gather}
	\ddot{y} + \dfrac{\gamma}{\mu} \dot{y} - y^{\prime \prime} = Mq, \label{ad_model_periodique_1} \\
	\ddot{q} + \varepsilon \left( q^{2}-1 \right) \dot{q} + q = A \ddot{y}, \label{ad_model_periodique_2}
	\end{gather}
\end{subequations}
with the condition
\begin{equation}
	y^{\prime} \left( 0,t \right) - y^{\prime} \left( l,t \right) = l \xi \dot{y} \left( l,t \right),
	\label{ad_BC}
\end{equation}
where $l = L \omega_{f} \sqrt{m_{t}/\Theta}$ is the dimensionless distance between dampers, $\xi = R/L m_{t} \omega_{f}$ is the equivalent damping coefficient per unit length and all other dimensionless variables are chosen as in the other sections. In this case, the efficiency reads
\begin{equation}
	\eta = \dfrac{\left\langle R \left( \dfrac{\partial Y}{\partial T} \right)^{2} \right\rangle}{\frac{1}{2} \rho D U^{3} L}\cdot
	\label{efficiency_formula_periodique}
\end{equation}
Using the variables defined above, the efficiency also reads $\eta = 16 \mu \pi^{3} S_{t}^{3} \left\langle \xi \dot{y}^{2} \right\rangle$, as in the previous cases. 

Equations \eqref{ad_model_periodique} and \eqref{ad_BC} are now integrated in time and space, using the same method as in the previous section, and the steady state dynamics of the cable is analyzed in terms of efficiency $\eta$. 

\subsection{Optimal energy harvesting}

\begin{figure}[!ht]
	\centering
	\psfrag{x0}[cc][br][0.75]{$10^{-2}$}
	\psfrag{x1}[cc][br][0.75]{$10^{-1}$}
	\psfrag{x2}[cc][bc][0.75]{$1$}
	\psfrag{x3}[cc][bc][0.75]{$10$}
	\psfrag{y1}[cc][lc][0.75]{$0$}
	\psfrag{y2}[cc][lc][0.75]{$0.1$}
	\psfrag{y3}[cc][lc][0.75]{$0.2$}	
	\psfrag{sigmadelta}[tc][bc][0.9][0]{$\xi$}
	\psfrag{eta}[rc][lc][0.9][-90]{$\eta$}	
	\includegraphics[width = 0.6\linewidth]{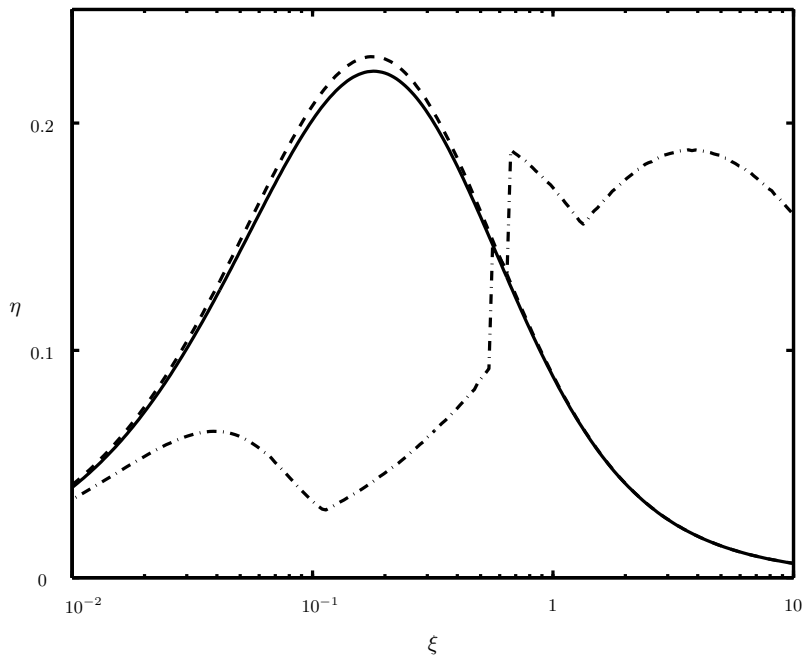} 
	\vspace{0.5cm}
	\caption{Efficiency as a function of the damping coefficient for the three considered configurations : the 2D case of an elastically mounted rigid cylinder (dashed), the infinite tensioned cable with continuous harvesting (solid) and the infinite tensioned cable with localized harvesting devices (dash-dot).}
	\label{Comparison_efficiency} 
\end{figure}

Two dimensionless parameters, $\xi$ and $l$, now govern the dynamics of the system. In order to compare with the results of the other geometries, we seek, for any given damping $\xi$, the maximum efficiency over all possible length $l$. Figure \ref{Comparison_efficiency} shows the comparison between the 3D-periodic case, the 3D-continuous case and the 2D case. It appears that (i) the maximum efficiency for the periodic harvesting, $\eta = 0.19$, is almost as high as for the other cases, (ii) the corresponding optimal damping per unit length is much higher than in the previous cases, $\xi = 3.65$, and (iii) the evolution of the efficiency with damping does not follow the classical bell shape. A more detailed analysis of the efficiency map is thus needed to understand these features.

\begin{figure}[!ht]
	\centering
	\psfrag{x0}[cc][br][0.75]{$10^{-2}$}
	\psfrag{x1}[cc][br][0.75]{$10^{-1}$}
	\psfrag{x2}[cc][bc][0.75]{$1$}
	\psfrag{x3}[cc][bc][0.75]{$10$}
	\psfrag{y1}[cc][cc][0.75]{$1$}
	\psfrag{y3}[cc][cc][0.75]{$3$}
	\psfrag{y5}[cc][cc][0.75]{$5$}
	\psfrag{y7}[cc][cc][0.75]{$7$}
	\psfrag{c1}[cc][rc][0.75]{$0.1$}
	\psfrag{c2}[cc][rc][0.75]{$0.2$}
	\psfrag{A}[cc][cc][0.9]{A}
	\psfrag{B}[cc][cc][0.9]{B}
	\psfrag{C}[cc][cc][0.9]{C}				
	\psfrag{xi}[tc][bc][1][0]{$\xi$}
	\psfrag{lreduit}[rc][lc][1][-90]{$\dfrac{l}{\pi}$}
	\psfrag{a}[cc][cc][0.75]{(a)}	
	\includegraphics[width = 0.6\linewidth]{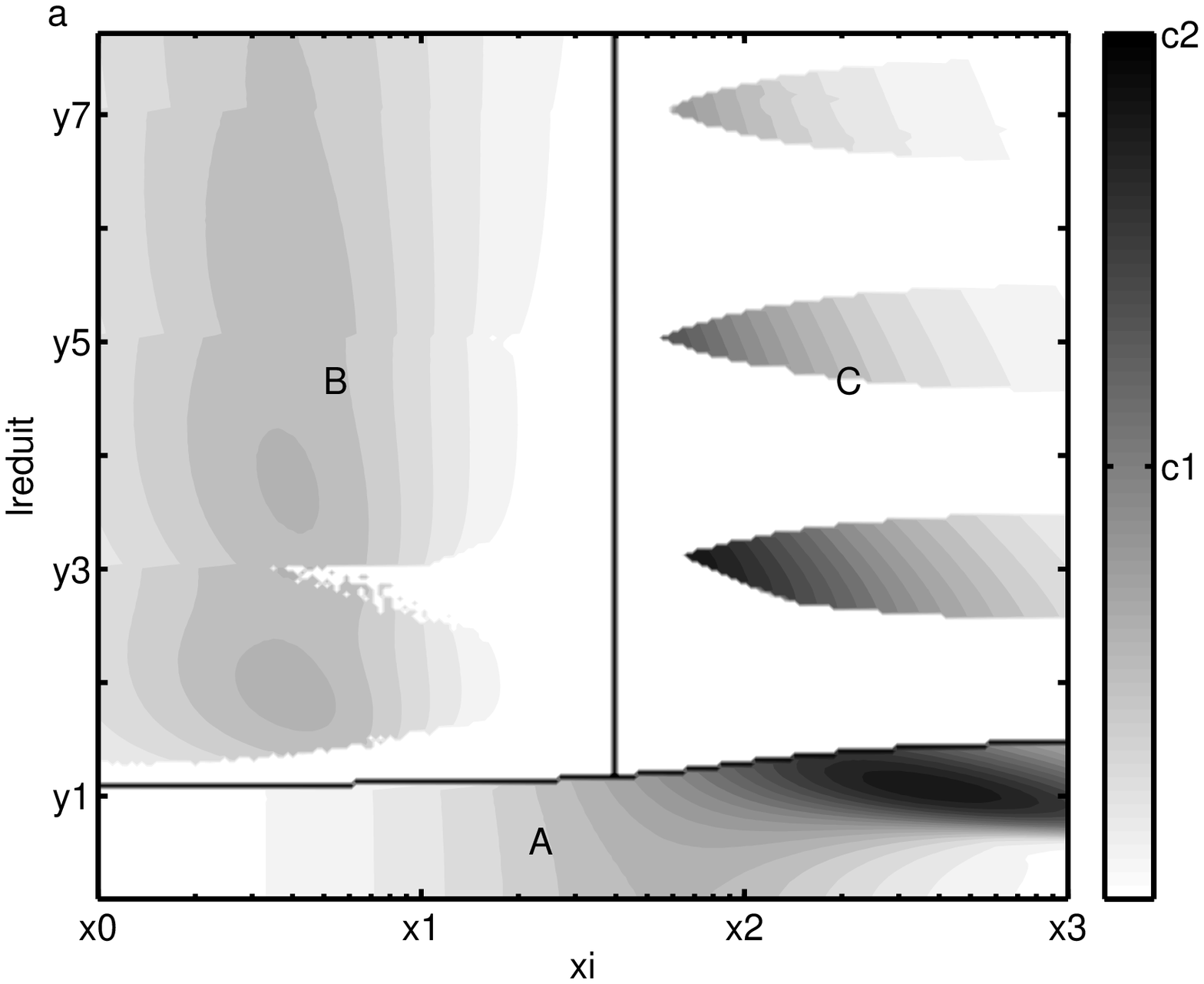} \vfill
	\vspace{0.5cm}
	\psfrag{x0}[cc][br][0.75]{$10^{-2}$}
	\psfrag{x1}[cc][br][0.75]{$10^{-1}$}
	\psfrag{x2}[cc][bc][0.75]{$1$}
	\psfrag{x3}[cc][bc][0.75]{$10$}	
	\psfrag{y1}[cc][lc][0.75]{$0$}
	\psfrag{y3}[cc][lc][0.75]{$0.2$}	
	\psfrag{A}[cc][cc][0.8]{A}
	\psfrag{B}[cc][cc][0.8]{B}
	\psfrag{C}[cc][cc][0.8]{C}		
	\psfrag{Xi}[tc][bc][1][0]{$\xi$}
	\psfrag{eta}[rc][lc][1][-90]{$\eta$}	
	\psfrag{b}[cc][cc][0.75]{(b)}	
	\includegraphics[width = 0.64\linewidth]{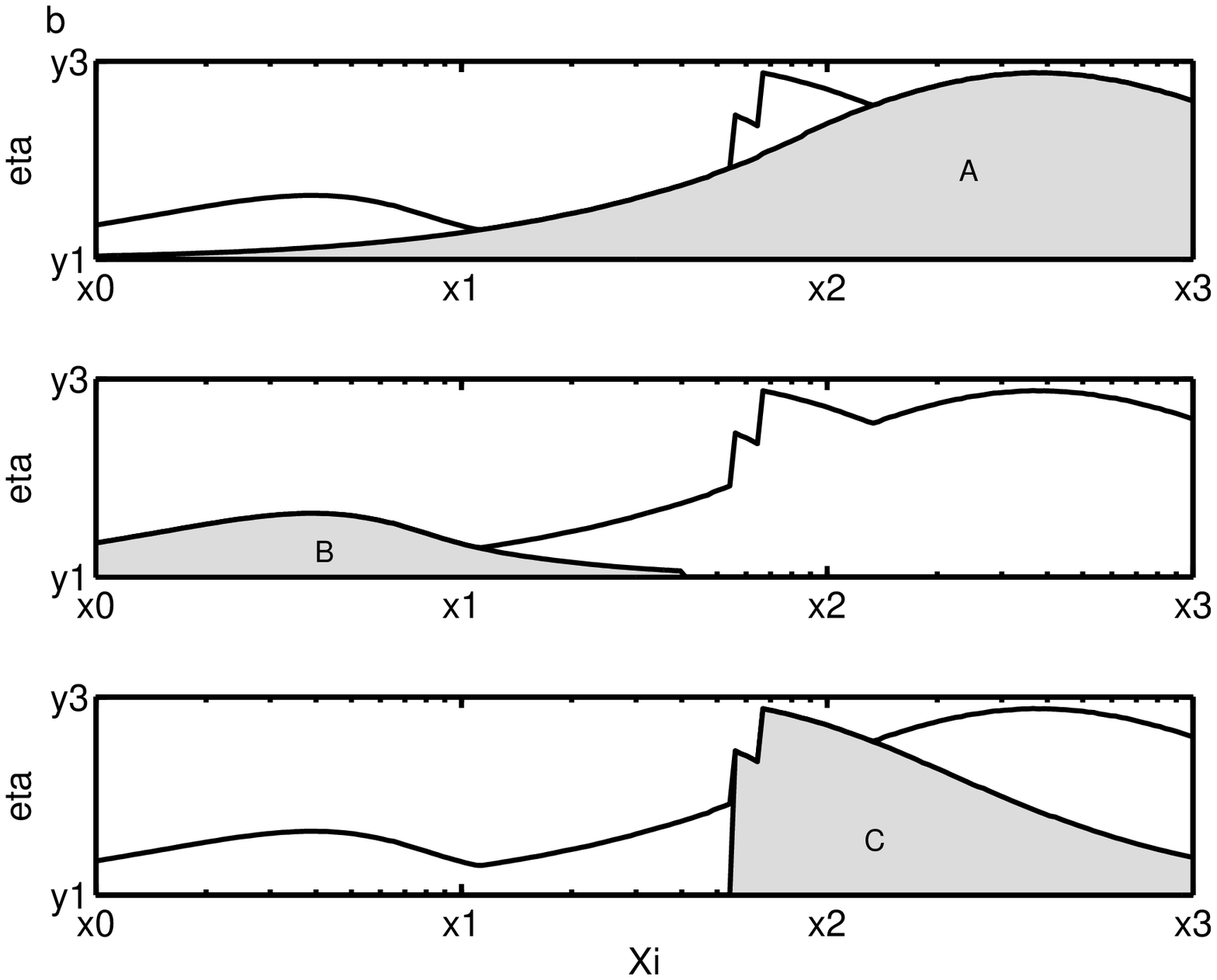}	
	\vspace{0.5cm}
	\caption{(a) Map of the energy harvesting efficiency from VIV of an infinite cable with discrete harvesting devices as a function of the damping density $\xi$ and the distance between two dashpots $l$ and clear identification of the three zones A,B and C of the parameters plane and (b) contribution of each zone in the efficiency curve of this configuration of energy harvesting.}
	\label{3D_periodique_efficiency} 
\end{figure}

The full map of efficiency is given in Figure \ref{3D_periodique_efficiency}(a). The effect of the parameters is complex, and may be analyzed by considering three zones. First, at the bottom of the map, or for small distances between harvesting devices (zone A) a simple effect of the damping coefficient is recovered : the evolution of the efficiency with the damping coefficient is close to the expected bell shape. The global maximum of efficiency also lies in this zone. In Figure \ref{3D_periodique_efficiency}(b), we also show that a large part of the full efficiency curve is explained by the variations of efficiency with the damping in this zone. For larger length but low damping, typically $\xi < 0.4$ (zone B), the efficiency is significantly smaller and only weakly depends on the length $l$. The evolution with $\xi$ is bell-shaped and the maximum over $l$, in this zone, corresponds to the bottom left part of the full efficiency curve (Figure \ref{3D_periodique_efficiency}b). A much more complex behaviour is found in the third zone (zone C). The efficiency strongly depends on $l$ : tongues of high efficiency are surrounded with negligible harvesting regions. These tongues are responsible for the discontinuities in the full efficiency curve as shown on Figure \ref{3D_periodique_efficiency}(b).

\begin{figure}[!ht]
	\centering
	\psfrag{x0}[tc][cc][0.65]{$-1$}
	\psfrag{x1}[tc][cc][0.65]{$0$}
	\psfrag{x2}[tc][cc][0.65]{$1$}
	\psfrag{y1}[cc][lc][0.65]{$0$}
	\psfrag{y2}[cc][lc][0.65]{$1$}		
	\psfrag{y3}[cc][lc][0.65]{$2$}
	\psfrag{y}[tc][bc][0.9][0]{$ y \left( z,t \right)$}
	\psfrag{z}[rc][lc][1][-90]{$\dfrac{z}{l}$}		
	\psfrag{a}[cc][cc][0.65][0]{(a)}	
	\psfrag{b}[cc][cc][0.65][0]{(b)}	
	\psfrag{c}[cc][cc][0.65][0]{(c)}	
	\psfrag{d}[cc][cc][0.65][0]{(d)}	
	\psfrag{e}[cc][cc][0.65][0]{(e)}	
	\includegraphics[width = 0.18\linewidth, height = 6.5cm]{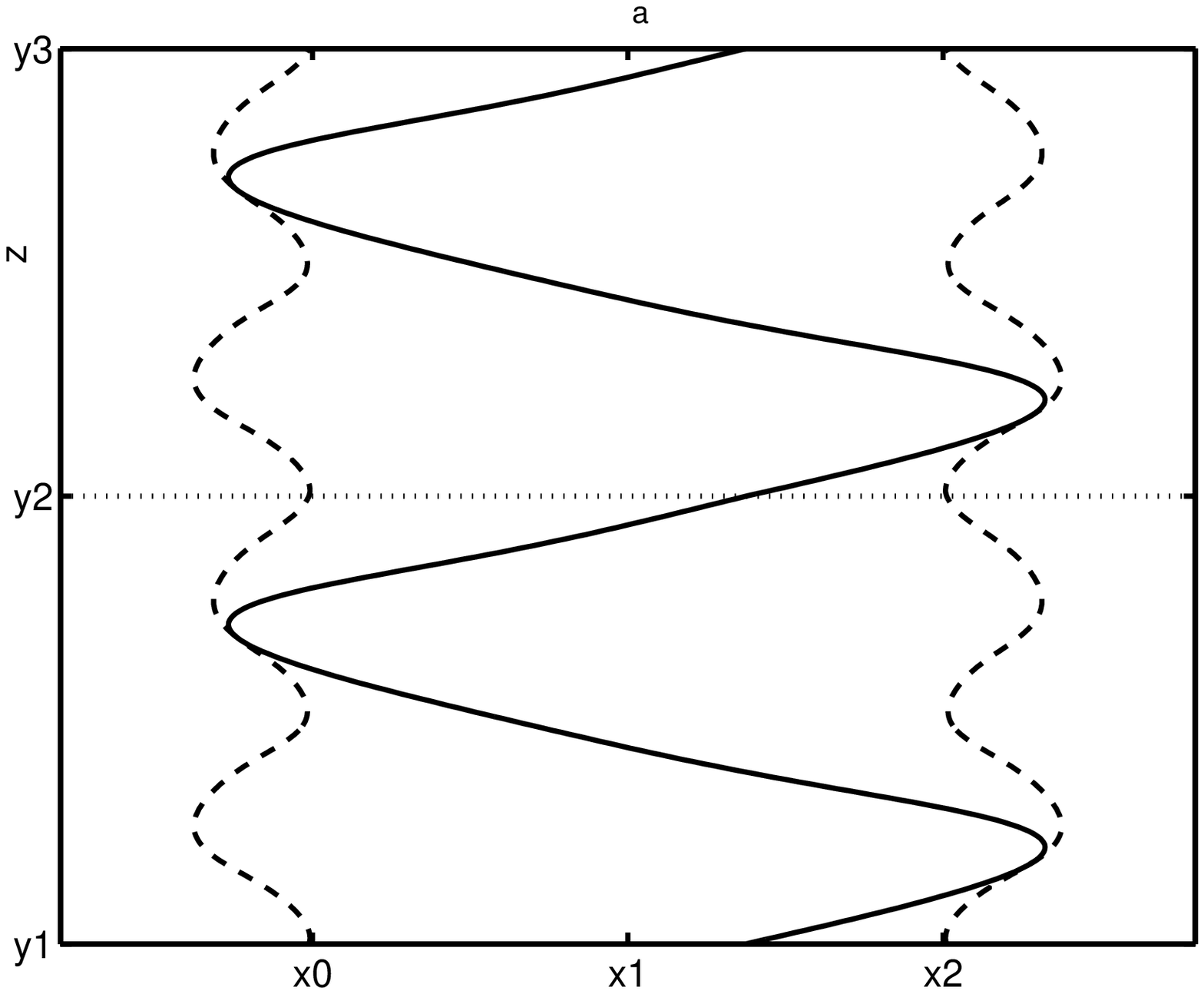} \hfill
	\includegraphics[width = 0.18\linewidth, height = 6.5cm]{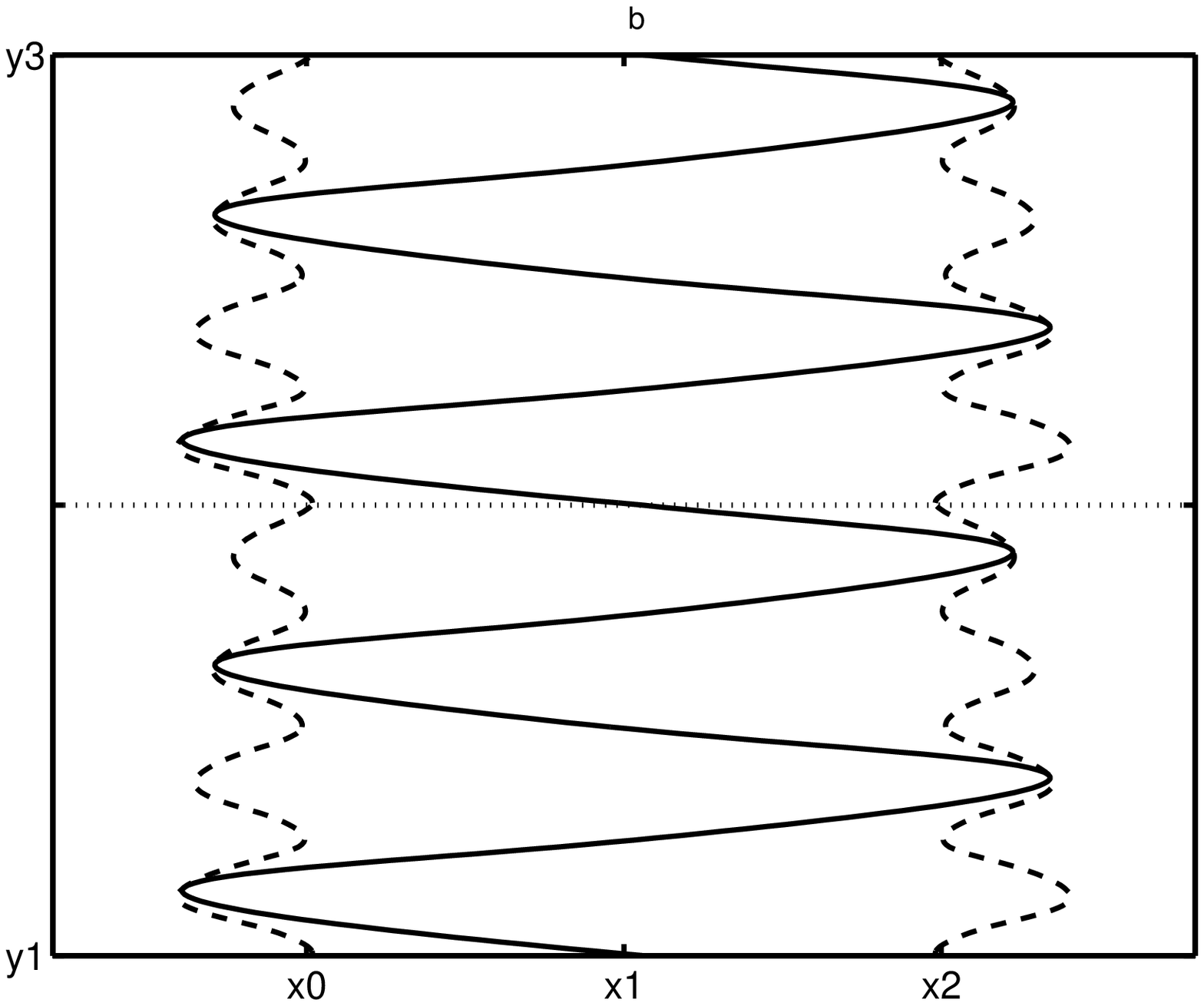} \hfill
	\includegraphics[width = 0.18\linewidth, height = 6.5cm]{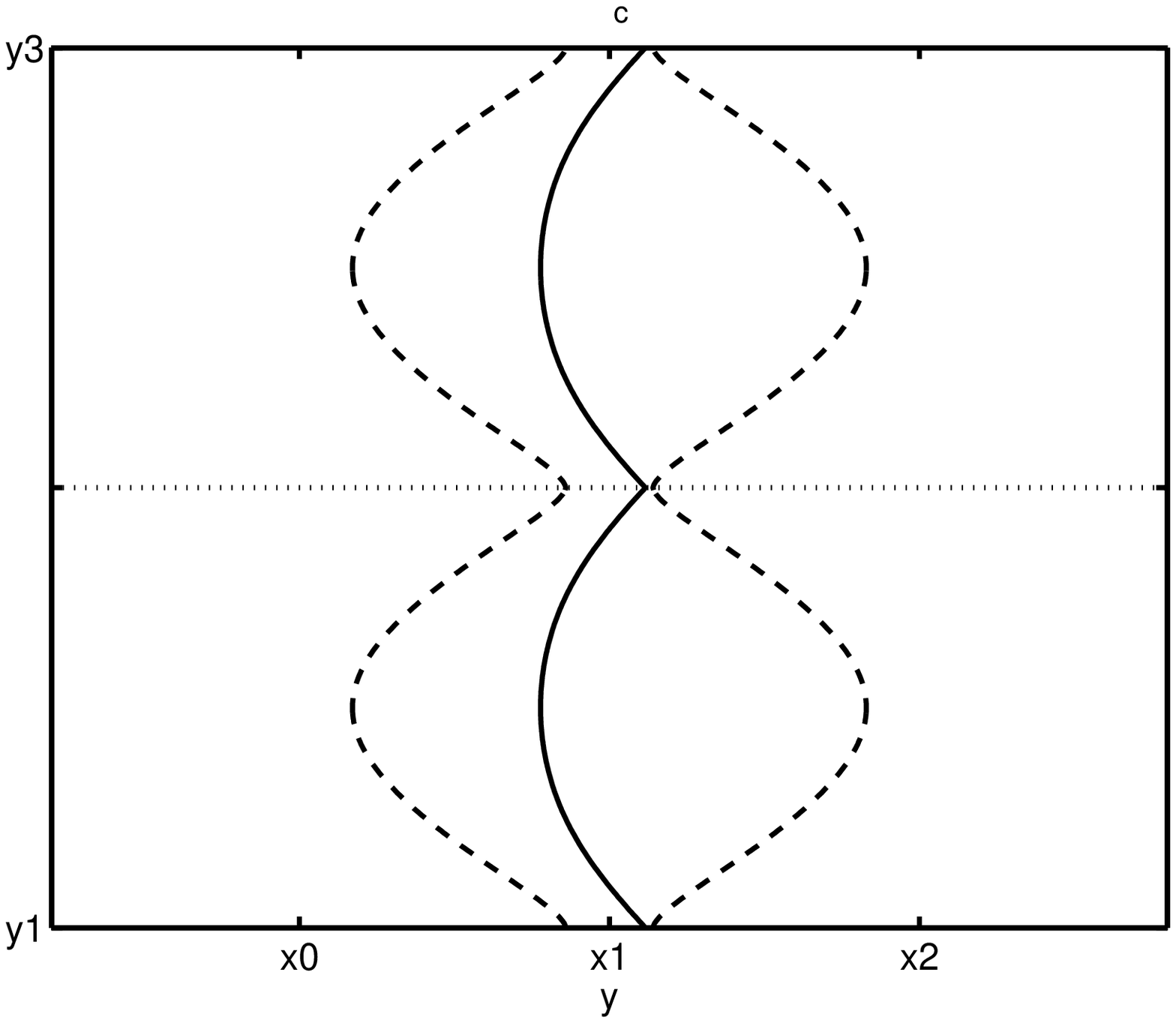} \hfill
	\includegraphics[width = 0.18\linewidth, height = 6.5cm]{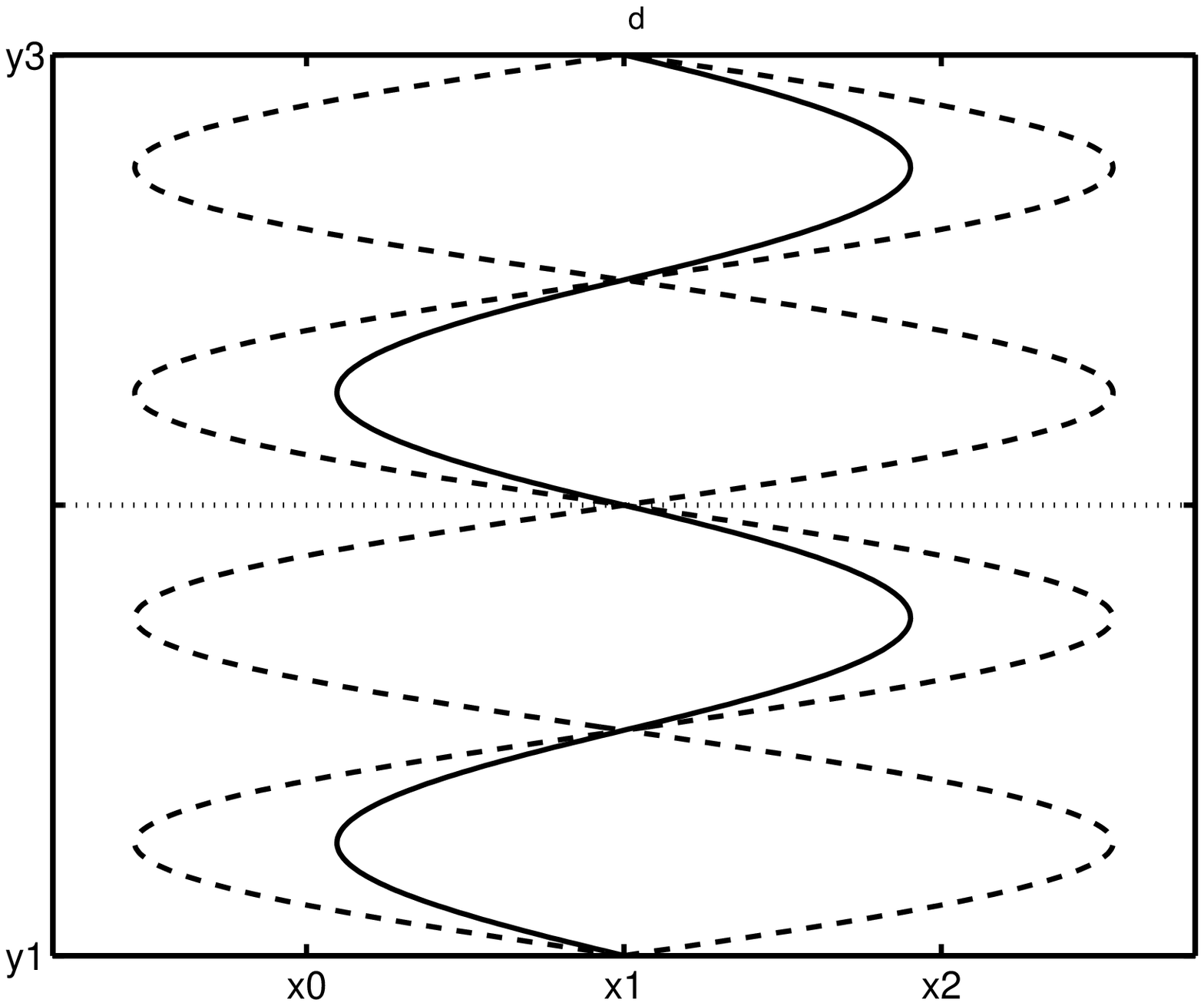} \hfill
	\includegraphics[width = 0.18\linewidth, height = 6.5cm]{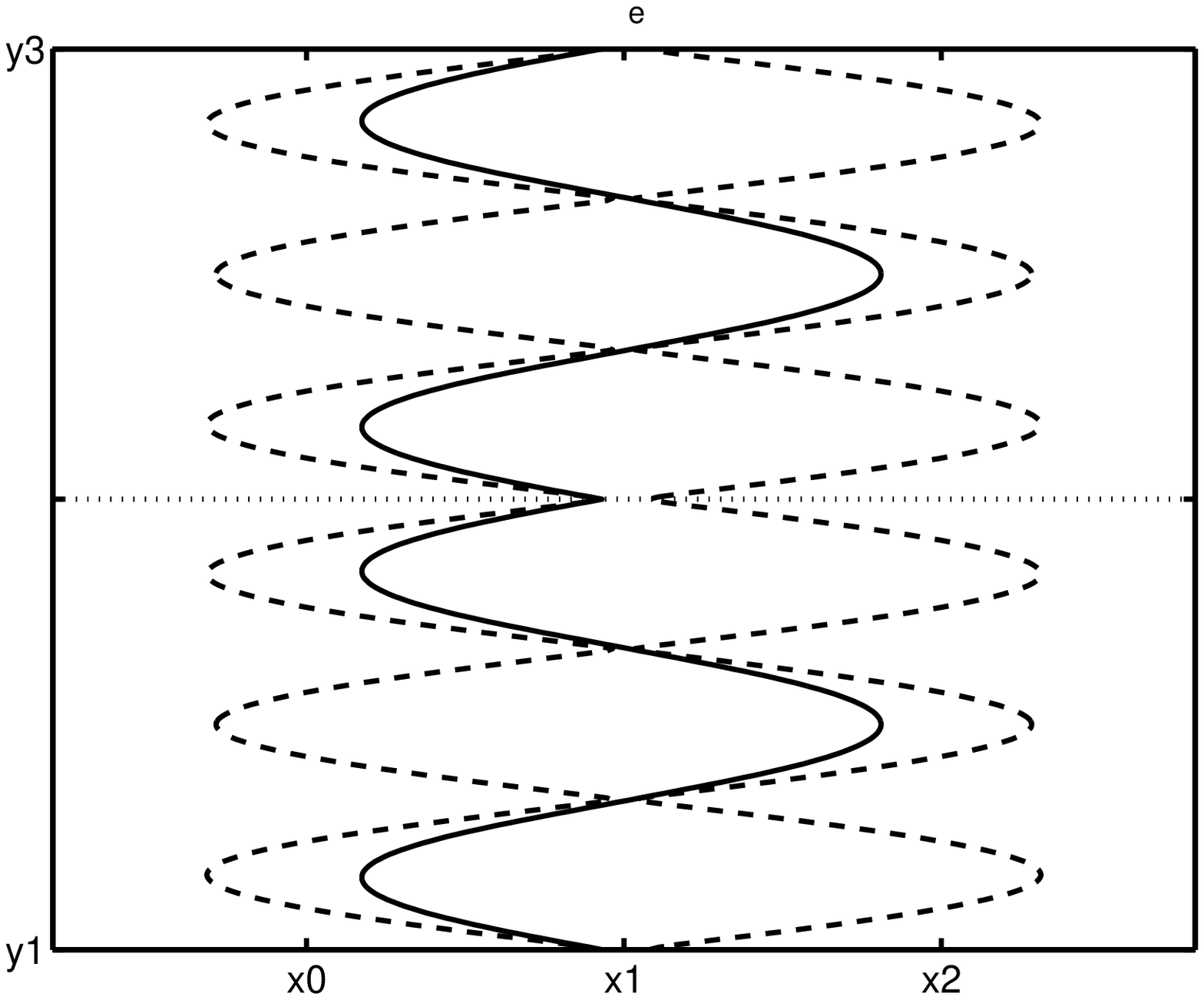} \hfill
	\vspace{0.5cm}
	\caption{Dynamics of the cable in different parts of the efficiency map, the solid line stands for the instantaneous position of the cable, the dashed line for the envelope of its motion and the horizontal dotted line recalls the position of one of the harvesters. (a) $l = 6.80$, $\xi = 0.0193$, $\eta = 0.05$, zone B, (b) $l = 13.41$, $\xi = 0.0193$, $\eta = 0.05$, zone B, (c) $l = 3.43$, $\xi = 3.65$, $\eta = 0.19$, zone A corresponding with the optimal case, (d) $l = 6.80$, $\xi = 3.65$, $\eta = 9.10^{-7}$, zone C, (e) $l = 9.56$, $\xi = 3.65$, $\eta = 0.07$, zone C.}
	\label{exemple_dynamiques}
\end{figure}

These zones differ essentially by the kinematics of the cable. A particular case is that of zone B, where the cable deforms in the shape of propagating waves, of wavelength $2 \pi$, $4 \pi$, $6 \pi$, ... depending on the length $l$. This is illustrated in Figure \ref{exemple_dynamiques}(a,b). As the efficiency is low in this zone, we shall not discuss it further. The optimal motion, which lies in zone A, is shown Figure \ref{exemple_dynamiques}(c). It is very close to the classical sinusoidal mode 1 of a tensioned cable, but with a small displacement of the support, which is responsible for the energy harvesting. All motions in zone A are of this type. At the same value of $\xi$ but for larger distances $l$, two cases relevant to zone C are displayed in Figure \ref{exemple_dynamiques}(d) and (e). For the first one, Figure \ref{exemple_dynamiques}(d), the efficiency is almost zero and the motion is similar to a mode 2 of a cable, with hardly any motion of the support. Conversely, Figure \ref{exemple_dynamiques}(e), a high efficiency is achieved with a motion resembling a mode 3, with support displacement.  

\subsection{Impact of the mode shapes on efficiency}

\begin{figure}[!ht]
	\centering
	\psfrag{impair}[tc][bc][.9][0]{odd mode}
	\psfrag{pair}[tc][bc][.9][0]{even mode}	
	\hspace{2cm}	
	\includegraphics[width = 0.2\linewidth]{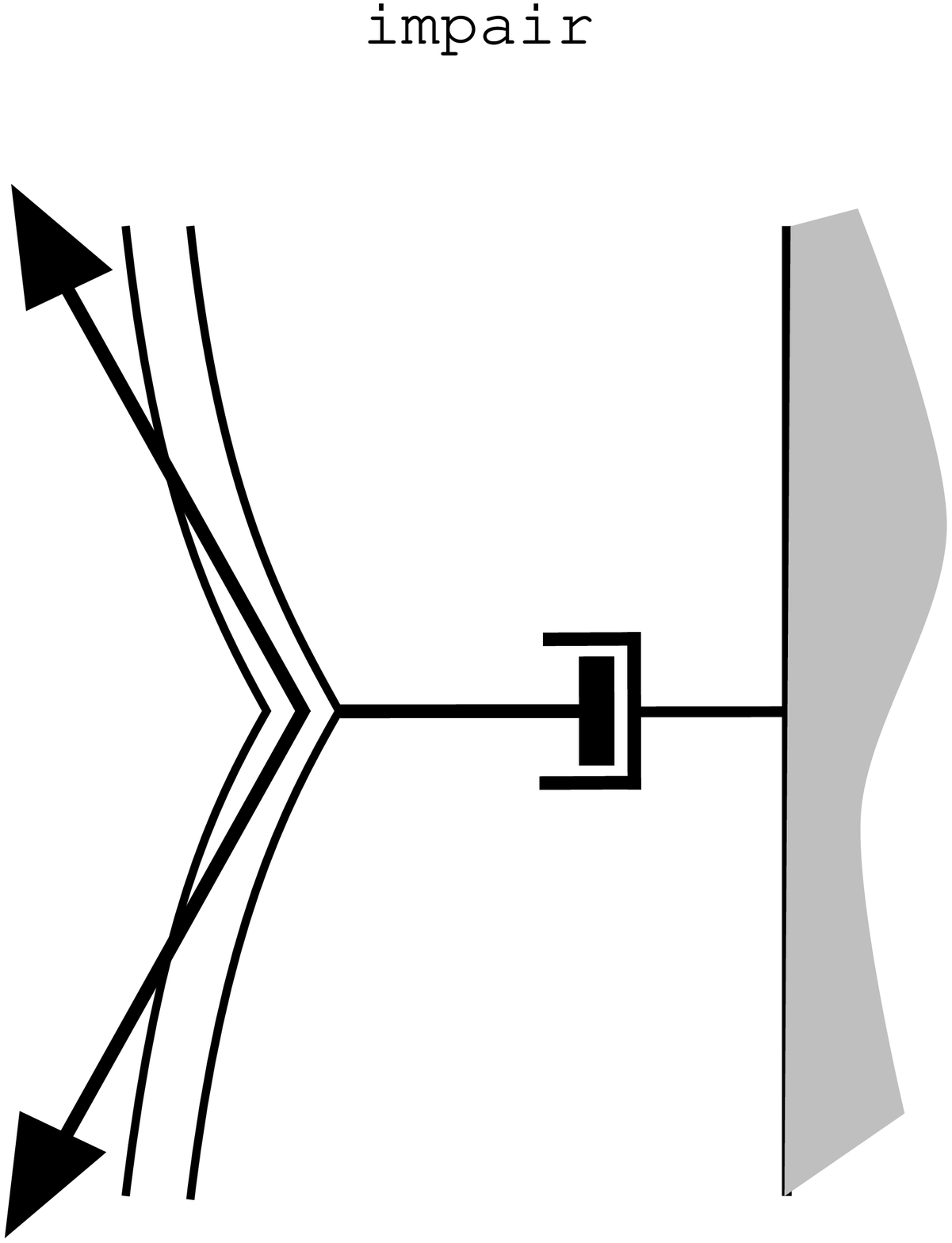} \hfill
	\includegraphics[width = 0.2\linewidth]{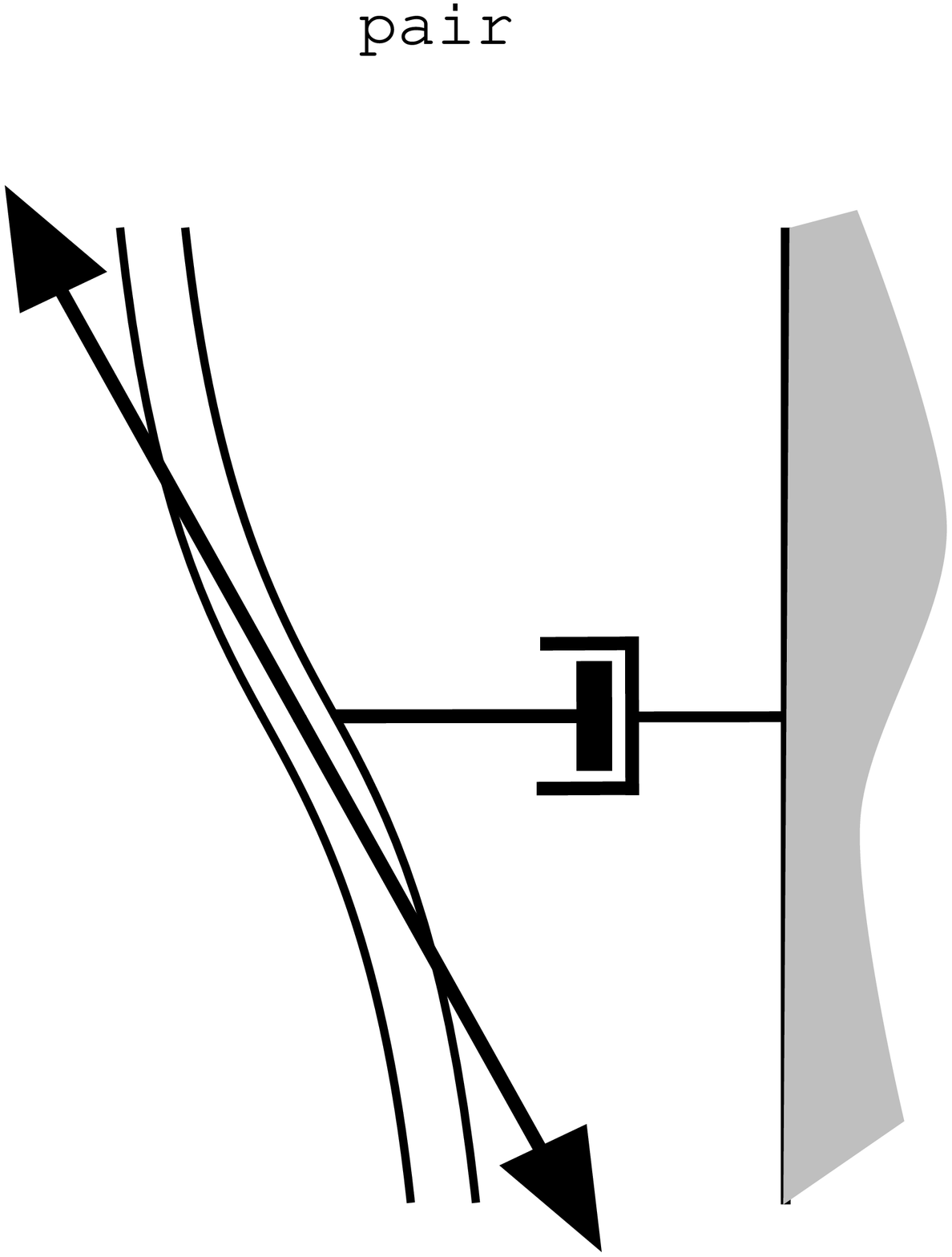}
	\hspace{2cm}
	\vspace{0.5cm}
	\caption{Mechanism of excitation of the harvester by stationnary modes, the arrows represent the local tension. Left : odd modes, high efficiency. Right : even modes, low efficiency.}
	\label{schemas_modes}
\end{figure}

Based on the comparison between the last three cases, we may expect that any motion resembling an odd mode of a cable will have a large efficiency, while even modes, which correspond to negligible forcing on the dampers, leads to inefficient harvesting. This is illustrated on Figure \ref{schemas_modes}, where the odd mode numbers will naturally exhibit a jump in slope, and thereby a force on the dashpot.

\begin{figure}[!ht]
	\centering
	\psfrag{x0}[cc][br][0.75]{$10^{-2}$}
	\psfrag{x1}[cc][br][0.75]{$10^{-1}$}
	\psfrag{x2}[cc][bc][0.75]{$1$}
	\psfrag{x3}[cc][bc][0.75]{$10$}
	\psfrag{y1}[cc][cc][0.75]{$1$}
	\psfrag{y3}[cc][cc][0.75]{$3$}
	\psfrag{y5}[cc][cc][0.75]{$5$}
	\psfrag{y7}[cc][cc][0.75]{$7$}
	\psfrag{c0}[lc][cc][0.75]{$0$}	
	\psfrag{c1}[lc][cc][0.75]{$0.1$}
	\psfrag{c2}[lc][cc][0.75]{$0.2$}		
	\psfrag{01}[lc][cc][0.95]{$0 \rightarrow 1$}
	\psfrag{2}[lc][cc][0.95]{$2$}
	\psfrag{3}[lc][cc][0.95]{$3$}
	\psfrag{4}[lc][cc][0.95]{$4$}
	\psfrag{5}[lc][cc][0.95]{$5$}
	\psfrag{6}[lc][cc][0.95]{$6$}
	\psfrag{7}[lc][cc][0.95]{$7$}
	\psfrag{8}[lc][cc][0.95]{$8$}
	\psfrag{xi}[tc][bc][1][0]{$\xi$}
	\psfrag{lreduit}[rc][lc][1][-90]{$\dfrac{l}{\pi}$}		
	\includegraphics[width = 0.6\linewidth]{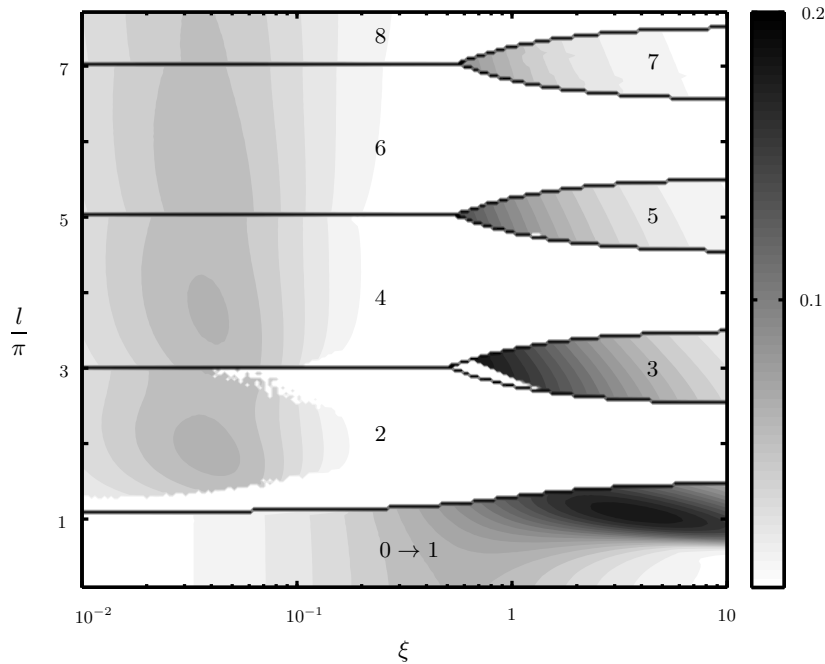} 
	\vspace{0.5cm}
	\caption{Correspondence between the mode numbers zones delimited by the solid lines and the efficiency map. Note that the odd mode numbers regions match very well with the high efficiency zones.}
	\label{numero_mode}
\end{figure}

As the mode number seems to be of primary importance in determining the efficiency, the linearized version of the equations is used to estimate it, as in previous sections. For each set of parameters $\left( \xi, l \right)$, the most unstable mode is analyzed in terms of its wavelength $\lambda$ and a corresponding mode number is defined as $n = 2 l / \lambda$. Note that this number may not be exactly an integer, due to the boundary condition : in that case the closest integer is used. The regions corresponding to each mode number are reported on the efficiency map (Figure \ref{numero_mode}). The regions of high efficiency in zone A and the tongues of zone C are all associated with odd mode numbers, as expected. 

In zone B, only even modes exist. Odd modes, which are the efficient ones, only appear for high damping, in zone A and C. This explains why the optimal efficiency of this configuration is obtained for rather high damping. Indeed, damping plays a double role here by transforming the overall mode shape and by controlling the local amplitude at the dampers locations. 

\section{Discussion}
\label{sectiondiscussion}

We may now compare the results of the three configurations analyzed above. The first feature to point out is that the optimal efficiencies are comparable, of the order of $0.2$. They are achieved with different values of damping density $\xi$, but always with a lock-in condition between the wake frequency and a frequency of the solid. These lock-in conditions deserve to be discussed further. In the simple 2D case, the lock-in condition corresponds to the coincidence between the wake frequency, $\omega_{f}$, and the natural frequency of the elastically supported body, $\omega_{s}$. It reads simply $\delta = \omega_{s}/\omega_{f} = 1$. In the second case, Section 3, the tensioned cable with continuous damping support has the ability to adapt its wavenumber $k$ so that the local frequency of motion synchronizes with the wake frequency, namely $k = 1$. When another length scale is introduced by separating the dampers with a distance $L$, the choice of the wavenumber is restricted, and lock-in is possible for each mode. Yet, harvesting will be significant only for odd modes, and they exist only for high enough damping, Figure \ref{numero_mode}. The lock-in condition reads, for these particular modes, $l = \left( 2 n + 1 \right) \pi$. 

Although lock-in, as described above, is an essential feature of an efficient harvesting, the optimal conditions differ slightly from exact lock-in. We now discuss this in terms of dimensional parameters. The optimal regime for the 2D case corresponds to
\begin{equation}
\delta = \left( \dfrac{D}{2 \pi S_{t} U} \right) \omega_{s} = 0.89, \hspace{1cm} \xi = \dfrac{r D}{2 \pi S_{t} m_{t} U} = 0.18, 
\label{2D_dim}
\end{equation}
and for the tensioned cable with localized harvesting devices
\begin{equation}
\dfrac{\pi}{l} = \left( \dfrac{D}{2 \pi S_{t} U} \right) \dfrac{\pi}{L} \sqrt{\dfrac{\Theta}{m_{t}}} = 0.92, \hspace{1cm} \xi = \dfrac{\left( R/L \right) D}{2 \pi S_{t} m_{t} U} = 3.65. 
\label{3D_dim}
\end{equation}
These show that they are of identical forms, except for the replacement of the rigid cylinder natural frequency $\omega_{s}$ in \eqref{2D_dim} by the cable mode frequency 
\begin{equation}
\omega_{c} = \dfrac{\pi}{L}\sqrt{\dfrac{\Theta}{m_{t}}}.
\label{puls_cable}
\end{equation}
For a given mechanical system, the optimal efficiency is reached only at one flow velocity. Yet, an advantage of the tensioned system is that the tension is easier to adapt than a mechanical elastic support stiffness. Noticeably, if the tension is caused by the flow, so that $\Theta \sim U^{2} $, the condition \eqref{3D_dim} over the distance between two dampers, $\pi / l = 0.92$ i.e. $l=3.43$, becomes independent of the velocity. As the velocity varies, the damping $\xi$ may shift, but it has a small influence, so that a near optimal efficiency is maintained. 

Using equation \eqref{efficiency_formula_periodique} and a value of $\eta = 0.18$, the power harvested by a cable of total length $L_{tot} = 100$ m, diameter $D = 0.04$ m and a $ 1.5$ m/s flow is about $1215$ W. If the tension comes from the flow, $\Theta = \rho D L_{tot} U^{2}$ and $\mu = 2$, as in \citet{Moda}, the dimensional optimal distance between two dampers is about $ L = 4$ m and twenty-five dampers are required along the cable to harvest those $1215$ W. 

All these results have been obtained under assumptions which shall now be discussed. A first assumption, used throughout the paper, is that motion is purely cross-flow, although VIV dynamics of a cable in flow are known to combine both in-line and cross-flow motions \citep{Vand}. Still, in-line motion amplitudes are generally smaller \citep{Huera}, and the harvesters we consider are purely cross-flow oriented. Assumptions on the cable are also made, by considering it as initially straight, and by neglecting non-linear geometrical effects in its dynamics. The wavelength of motion are here much larger than the cable diameter, discarding non-linear effects, and tension is such that the natural radius of curvature is much larger than the same wavelength. This was the case of cable systems such as in \citet{Viol} or \citet{Xu}. The fundamental model used throughout the paper is that of a wake oscillator. This approach has some limitations, particularly in describing the fine dynamics of the wake, as those computed for instance in \citet{Bour}, but it is known to capture well features such as lock-in and generic parameters dependences. As all the results given in this paper reduce to simple considerations on those aspects, we expect that they are, at least qualitatively, relevant and general enough. 

As a conclusion, we may state that energy harvesting using vortex-induced vibrations seems as feasible and promising using tensioned cables as it is using the classical rigid cylinder geometry. It may even have some specific advantages such as the ability to harvest energy with large devices, or the ability to permanently adapt the cable dynamics to the lock-in condition.

\end{document}